\documentclass[numbers,times,5p]{elsarticle}

\pdfoutput=1
\graphicspath{{./articlebreitung/breitungfigures/}}
\renewcommand{\citet}{\citep} 
\usepackage[utf8]{inputenc}
\usepackage[T1]{fontenc}
\usepackage[english]{babel}
\usepackage{caption}
\usepackage{subcaption}
\usepackage{graphicx}
\usepackage{psfrag}
\usepackage[usenames]{color}
\usepackage{exscale}
\usepackage{ifthen}
\usepackage{fancyhdr}
\usepackage{amssymb} 
\usepackage{amsmath}
\usepackage{mathtools}
\usepackage{url}

\newcommand{\ba}{\begin{eqnarray}}
\newcommand{\ea}{\end{eqnarray}}

\newcommand{\bann}{\begin{eqnarray*}}
	\newcommand{\eann}{\end{eqnarray*}}

\newcommand{\bb}{\begin{Beweis}}
	\newcommand{\eb}{\end{Beweis}}

\newcommand{\be}{\begin{equation}}
\newcommand{\ee}{\end{equation}}

\newcommand{\bex}{\begin{bei}}
	\newcommand{\eex}{\end{bei}}

	\newcommand{\ein}{\frac{1}{2}}

	\renewcommand{\Pr}{\hbox{P}}
	\newcommand{\Er}{\hbox{$I\!\!E$}}

	\newcommand{\ul}[1]{\mbox{\boldmath  $ #1$}}
	\newcommand{\corr}{\mbox{\rm corr}}
	\newcommand{\cov}{\mbox{\rm cov}}

	\newcommand{\var}{\mbox{\rm var}}
	
	\newcommand{\uH}{\ul{H}}
	
	\newcommand{\uI}{\ul{I}}

	\newcommand{\uP}{\ul{P}}

	\newcommand{\uU}{\ul{U}}
	\newcommand{\uu}{\ul{u}}

	\newcommand{\ux}{\ul{x}}
	\newcommand{\uX}{\ul{X}}

\begin{document}

	\bibliographystyle{../elsarticle/elsarticle-num}
		
		
	\begin{frontmatter}
		
		\title{	The Geometry of Limit State Function Graphs and Subset Simulation}
		\author[mymainaddress]{Karl Breitung \corref{mycorrespondingauthor}}
		\address[mymainaddress]{Engineering Risk Analysis Group,    Technical University of Munich,  Theresienstr.~90, 80333 Munich, Germany}
		\cortext[mycorrespondingauthor]{Corresponding author}
		\ead{breitu@aol.com}
				
		
		
		\begin{abstract}
			In the last fifteen  the subset sampling method has often  been used  in reliability problems as a tool for calculating  small probabilities. This method is extrapolating  from an initial Monte Carlo estimate for the probability content of a failure domain found by a suitable higher level of the original limit state function. Then iteratively conditional probabilities are estimated for  failures domains decreasing to the original failure domain. 	
			But  there are   assumptions not immediately obvious  about the structure of the failure domains which must  be fulfilled that the method works properly.  Here   examples are studied that show that at least in some cases if these premises are not fulfilled, inaccurate results may be obtained. For the further development of the subset sampling method it is certainly  desirable to find approaches where it is possible to check that these implicit assumptions are not violated. Also it would be probably important to develop further improvements of the concept to get rid of these limitations.
					\end{abstract}
		
		
	\end{frontmatter}	
\thispagestyle{empty}
\section{Introduction} 
A standard problem of structural reliability is the calculation  of failure probabilities which are given by $n$-dimensional integrals in the form:
\ba
P=\int_{ g(\scriptsize{\ux)}<0 }f(\ux)\ d\ux
\ea
Here $f(\ux)$ is an $n$-dimensional PDF (probability density function)
and $g(\ux)$ is the LSF (limit state function) describing the failure condition. 
During the development of structural reliability methods  it became usual to transform the random vector $\uX$ into a standard normal
random vector $\uU$ with independents components. So the problem is in this standardized form:
\ba
P=(2\pi)^{-n/2}\int_{ g(\scriptsize{\uu})<0 }\exp(-\ein|\uu|^2)\ d\uu
\ea
Such  transformations into the the standard normal space for random vectors with independent components were first described by  \citet{Rackwitz/Fiessler(1978)}.
For  random vectors with dependent components the Rosenblatt-transformation is often proposed to achieve a transformation but with the exception of the example given in \citet{Hohenbichler(1981)} no applications of this transformation concept are known to the author. A practically applicable method   appears to be the Nataf-transformation described in \citet{Der/Liu(1986)}.

From   the sixties last century for this problem basically two different solution concepts  were followed.
The first were Monte-Carlo methods. The second  were the so-called FORM/SORM concepts.
In the further development then  hybrid methods  originated, combining both concepts. A review 
over these developments till end of the nineties can be found  in \citet{Rackwitz(2001)}.

During the decades the problems in structural reliability became more complex, i.e.   higher dimensional problems with smaller probabilities had to be solved. 
A new method for such tasks these was proposed in 2001, the so-called subset simulation method  (SuS) -- the author prefers this acronym -- will be described in the next section. A connection of this approach with FORM/SORM is described in section~3 and in the next section the approximation for the estimation variance used in SuS is explained in more detail. Then in  section~5 some examples using SuS will be studied  to show  that SuS has some implicit assumptions. In the case that these are not fulfilled the results obtained by SuS might be misleading.  
This paper is an enlarged and revised version of  the conference paper \citet{Breitung(2016)}.
\section{The Subset Simulation Concept}
The subset simulation method is basically a variant of Monte Carlo methods; it  attempts  to avoid the large quantity of data points which has to be created in standard Monte Carlo by using instead an iterative procedure.
It can be subsumed under the generic term of  stochastic optimization procedures.
Whereas for example importance sampling tries to improve the efficiency of Monte Carlo by identifying regions with high probability content and putting more date points  there,  SuS starts from a region around the origin and then moves step by step towards the   failure domain. These regions are defined here by domains in the form $F_i=\{g(\uu)< a_i \}$ with the $a_i$'s being positive and $a_i \to 0 $.

The basic thought of the method (see \citet{Au/Beck(2001)}, \citet{Au/Wang(2014)}) is now to write  the failure probability $\Pr(F)$ as a product of
conditional probabilities
\ba
\hspace*{-0.85cm}\Pr(F_n)=\Pr(F_1|F_0)\!\cdot\!\Pr(F_2|F_1)\ldots\Pr(F_n|F_{n-1}) = \prod_{k=0}^{n-1}\Pr(F_{k+1}|F_k)\hspace{-0.2cm}
\ea
Here  $\mathbb{R}^n=F_0\supset F_1 \supset F_2\supset \ldots \supset F_n=F$.
Since the  respective (suitably chosen) conditional probabilities are relatively  large compared with the failure probability $\Pr(F)$ which has to be estimated, such an access to the problem  has the advantage that these conditional probabilities can be estimated  more efficiently with much   smaller sample sizes. The details how these samples are produced with Monte Carlo Markov chains can be found in the references given above.
\begin{figure}[hbt]
	\begin{center}
		\includegraphics[width=8cm,height=5cm]{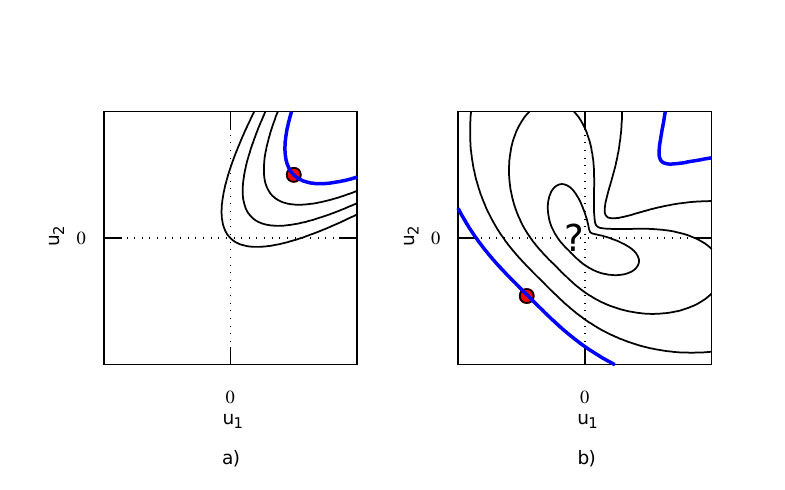}
		\caption{Figure a) the standard SuS example, figure b) a more complicated LSF. The design points  red, curve $g(u_1,u_2)=0$ blue.}
		\label{fig:gglsfbeian}
	\end{center}
\end{figure}
This construct is an iterated extrapolation starting from an initial probability estimate $\widehat{\Pr(F_1)}$ 
of a much larger failure domain. Then iteratively the failure domain is shrunk towards the original failure domain. In many papers about SuS the examples are similar the case shown in figure \ref{fig:gglsfbeian}a), a standard example for demonstrating the concept. But one should study also how  the method works for more complex cases as e.g. in figure \ref{fig:gglsfbeian}b).

As Rackwitz said in \citet{Rackwitz(2001)},  an crucial step in  developing  new  procedures  is also to show where they do not work, i.e. to  construct counterexamples. In practical mathematical methods almost never a proof can be found showing the correctness of an approach, but with examples one can show the limits of the applicability of the concept and where it needs improvement. On the other hand it is practically impossible to show the correctness of a method by some examples.

So to understand the thought of SuS more clearly, it is profitable to  study  examples where the concept runs into difficulties which will be done in section~5. This might either lead to an improvement of the used procedures which then allows to tackle also such examples successfully or to see possible limitations more distinctly. 
\section{SuS and Asymptotic Approximations}
\begin{figure}
	\begin{center}
\includegraphics[width=8cm,height=5cm]{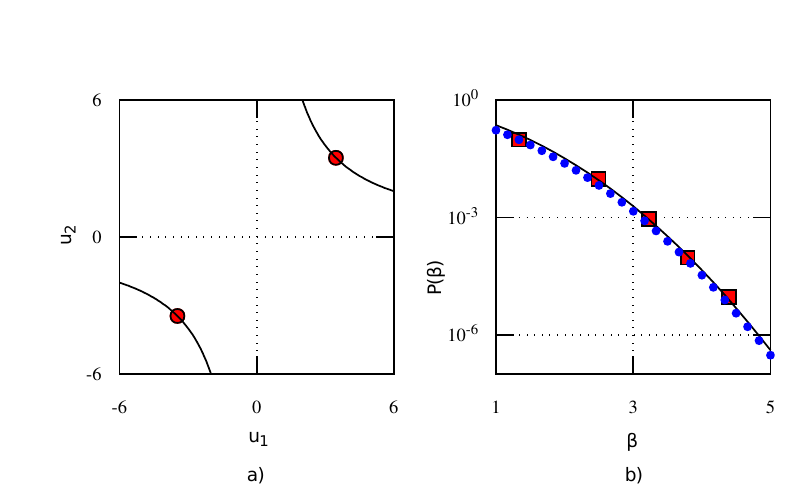}
\caption{Left: limit state surface (solid black line) for $\beta=\sqrt{12}$ and design points (red circles), right:  asymptotic approximation (solid  line) and  curve  (dotted blue  line) fitted from SuS data (red squares
	)}
\label{fig:sormfig}
\end{center}
\end{figure}

There are interesting relations between SORM and SuS which seem to have gone unnoticed till now. If one considers  the first example in chap.~5 in \citet{Au/Wang(2014)}, the 
 LSF's there are given by
\ba\label{lsf}
g_{\beta}(u_1,u_2)=\frac{\beta^2}{2}-u_1\cdot u_2.
\ea
Taking now the constant in the LSF as in eq.(\ref{lsf}) gives a sequence of limit state surfaces with
$\min_{g_{\beta}(\uu)=0}|\uu|=\beta$.
The probabilities $\Pr(g_{\beta}(u_1,u_2)<0)$ can be calculated exactly. This is not  mentioned in \citet{Au/Wang(2014)}, but  one  has for $\beta>0$ that (\citet{Weisstein(2016)}):
\ba\label{prod}
\Pr(g_{\beta}(u_1,u_2)<0)=\pi^{-1}K_0(\beta^2/2),
\ea
where $K_0=(.)$ is the modified Bessel function of the second kind and of order $0$. 
 For this  Bessel function one has the following asymptotic approximation (see \citet{Abramowitz}, eq.~9.7.2, p.~378):
 \ba
K_0(z) \sim \sqrt{\pi/z}\cdot e^{-z},\ z\to\infty
\ea 
Inserting this  into eq.~\ref{prod} and then using Mill's ratio 
(see \citet{Abramowitz}, eq.~26.2.12, p.~932)
yields 
\ba\label{exact}
\Pr(g_{\beta}(u_1,u_2)<0)\sim \sqrt{2}\cdot \Phi(-\beta), \ \beta\to\infty
\ea
With  the SORM approach now asymptotic approximations can be computed.
 Using the Lagrange multiplier method two beta points can be  found $\uu_1=(\beta/\sqrt{2},\beta/\sqrt{2}  )$ and $\uu_2=(-\beta/\sqrt{2},-\beta/\sqrt{2})$.
At these points one has ($\uI_n$ denoting the $n$-dimensional unit matrix):
\ba\nonumber 
\nabla g_{\beta}(\uu)=\left(\begin{array}{r} - \frac{\beta}{\sqrt{2}}\\  \\ -\frac{\beta}{\sqrt{2}}\end{array}\right)&,&\ 
\nabla^2 g_{\beta}(\uu)=\left(
\begin{array}{rr} 0 & -1\\-1& 0 \end{array}\right),\ 
\\ \nonumber
\uP= \left(
\begin{array}{rr} \ein & \ein \\ & \\ \ein  & \ein \end{array}\right)&,&\  
\uI_2-\uP=\left(
\begin{array}{rr} \ein & -\ein\\ & \\-\ein & \ein \end{array}\right),
\\ \tilde{\uH}&=& \ul{I}_2+\beta |\nabla_{\beta} g(\uu)|^{-1}\nabla^2 g_{\beta}(\uu)
\ea
The SORM approximation is found adding the equal contributions from the two beta points.
 As described in \citet{Breitung(2015)}, one obtains as SORM approximations then:
\ba
\hspace{-0.8cm}\Pr(g_{\beta}(u_1,u_2)\!<0)\!\sim\! \frac{2\!\cdot\! \Phi(-\beta)}{\sqrt{\det((\uI_2\!-\!\uP)\tilde{\uH}(\uI_2\!-\!\uP)+\uP)}},\hspace{0.05cm} \beta\to\infty
\ea
Inserting the values found before, one finds:
\ba
\Pr(g_{\beta}(u_1,u_2)<0)\sim \sqrt{2}\cdot\Phi(-\beta)
,\ \beta\to\infty
\ea
This agrees with the asymptotic approximation for the
exact probability  given in eq.~\ref{exact}. 
In  figure~\ref{fig:sormfig} now they 
are compared. 
In \citet{Breitung(1994b)}, chap.~6 it is explained that 
given a LSF such that $ \min_{g(\uu)=0}|\uu| =1$ 
and defining
\ba
F(\beta)=\{\uu; g(\beta \uu )< 0)\}, \ P(\beta)=P(F(\beta)),
\ea
one has for this sequence of failure domains  that
\ba
P(\beta)\sim c\beta^b \cdot \Phi(-\beta),\ \beta\to\infty
\ea
with $c,b\geq 0$ non-negative constants. 
Therefore the results of SuS can be used to estimate the constants in these asymptotic approximations with the proviso that these results are unbiased (see section~\ref{bias} for a discussion).

In figure \ref{fig:sormfig} from four values of the complimentary cumulative distribution function estimated by SuS (taken from \citet{Au/Wang(2014)}, p.~165, figure~5.3), marked by circles, the value of $c$ is estimated ($b$  is zero here) and the corresponding curve is drawn. It agrees quite well with the asymptotic approximation shown as  solid curve.

In the same way, for  stationary Gaussian vector processes $\ux(t)$   results can be found (see \citet{Breitung(1994b)}, chap.~$8$) for  outcrossing rates out of limit state surfaces. Here again SuS might be useful for determining the parameter values in the asymptotic formul\ae.
 \section{Bias and variance of SuS Estimates}\label{bias}
 In the derivation of  SuS some assumptions and approximations are made which are not spelt out too clearly.
 Let be given an estimate $\hat{P}$ of a failure probability $P$. Then the mean square error (MSE) 
 of the estimator is 
 \ba\label{MSE}
 \mathrm{MSE}(\hat{P})= \var(\hat{P})+(P-\Er(\hat{P}))^2
 \ea
 The first term on the rhs is the variance of the estimator and the second one its bias. 
 In the derivation of the SuS often  it is assumed  that the second term in  eq.~(\ref{MSE}) can be neglected.
 These assumptions is based on slightly cavalier arguments that somehow everything goes to infinity and therefore
 asymptotically large sample approximations can be used which state that the bias vanishes. Using asymptotic arguments is a problematic solution  for such questions, since the convergence speed to the theoretical asymptotic values cannot be estimated here by explicit error bounds.
 An argument against is further, that since SuS is an iterative method, the bias could accumulate and might be amplified in the further stages.
 Therefore such argumentations  should always be underpinned by extensive Monte-Carlo studies to demonstrate if in realistic settings the claimed effect is observed in fact. 
 As far as known to the author this has not yet be done, so this part of the SuS reasonings remains still slightly speculative.
 
In \citet{Au/Beck(2001)} , \citet{Au/Wang(2014)} and \citet{Papaioannou(2015)} relations between the coefficient of variation of the estimator $\widehat{\Pr(F)}$ and the conditional probability estimators $\widehat{\Pr(F_i|F_{i-1})}$ are derived in a slightly sketchy way. Here it is attempted to rewrite the basic thought in a more precise  
 way showing all intermediate steps. The notation is also changed; the usual mathematical notational conventions for probability and statistics are observed  instead of the slightly unusual notations in the SuS papers.
 
 The following notation will be used:
 $P$ is the unknown failure probability, $P_i=P(F_i|F_{i-1})$ are the conditional probabilities,
 the estimators for them are denoted by a  hat over the probability, i.e.   $\hat{P}_i$. The realizations of the random variables and constants are denoted by lowercase letters, e.g. the realizations of 
 $\hat{P}_i$ in a run are denoted by $\hat{p}_i$. The mean value of a rv $X$ is denoted by $\Er(X)$, its variance by $\var(X)$ and the covariance between two rv's $X$ and $Y$ by $\cov(X,Y)$ and their correlation by $\corr(X,Y)$.
 The coefficient of variation of a random variable $X$ is denoted by $c_v(X)=\sqrt{\var(X)}/\Er(X)$.
 
 The following approximations for moments of functions of random variables are  in some texts called the \textit{ delta method} (see \citet{Casella/Berger(2002)}, \cite{Oehlert(1992)}), whereas in   \cite{Papoulis(1965)} they are  derived in a section \textit{approximate evaluation of moments} without giving them a specific name.
 For a sufficiently smooth function $h$ of a random variable $X$ one has in the univariate case the following approximation for the first two moments of the random variable $h(X)$
 \ba
 \Er(h(X))&\approx &h(\Er(X))\\
 \var(h(X))&\approx & h'(\Er(X))^2\cdot  \var(X)
 \ea
In the same way for a random vector $\uX=(\uX_1,\ldots,\uX_n)$ and a smooth function $h$ of it  this has then the form
 \ba\label{multidelta}\nonumber
\hspace*{-0.5cm} \Er(h(\uX))&\approx& h(\Er(x_1,\ldots,x_n))=h(\Er(\uX))\\
\hspace*{-0.5cm} \var(h(\uX))&\approx& \sum_{i,j=1}^n h_i(\Er(\uX))\ h_j(\Er(\uX))\cdot \cov(X_i,X_j)
 \ea
 with $h_i(\ux)=\frac{\partial h(\ux)}{\partial x_i}$ the partial derivative of $h$ with respect to $x_i$.
 These approximation are exact only if the function $h$ is linear and the random variables are normal. Otherwise the approximation errors depend on the non-linearity of the functions, the form of the distributions and --- in the multivariate case --- the dependence between the components of $\uX$.
 Therefore information about the quality of this  
 approximation in SuS can be obtained only by numerical experiments which seem 
 to be still lacking.
 
An approximation for the variance of $\hat{P}$ can now found using  the delta method in its multivariate version in eq.~(\ref{multidelta}) and assuming $\Er(\hat{P}_i)=P_i$:
 \ba
 \hspace*{-0.8cm}\var(\hat{P})\approx \var\left( \prod_{i=1}^n \hat{P}_i \right) = \sum_{i,j=1}^n\left[ \prod_{k\neq i}P_k \prod_{l\neq j}P_l\right]\cov(\hat{P}_i,\hat{P}_j)
 \ea
 Dividing this by $\Er(\hat{P})^2$  and approximating it by $(\prod_{i=1}^n P_i)^2$ on the rhs one obtains:
 \ba
 \frac{\var(\hat{P})}{\Er(\hat{P})^2}\approx  \sum_{i,j=1}^n \frac{\cov(\hat{P}_i,\hat{P}_j)}{P_i\cdot P_j}
 \ea
 Now using the definition of the $c_v$ this can be written as
 \ba
 c_v(\hat{P})\approx  \sum_{i,j=1}^n c_v(\hat{P}_i)c_v(\hat{P}_j)\cdot \corr(\hat{P}_i,\hat{P}_j)
 \ea
 This equation is given in  \citet{Papaioannou(2015)}, but without derivation.
 The derivation here shows that some assumptions are involved.
 
%
 
 Now, in the papers about SuS the coefficient of variation is used for calculating some sort of confidence intervals.
  The problem is that for deriving a meaningful confidence interval, it is not enough to calculate the sample mean and sample standard deviation. Additionally the form of the distribution of the sample has to be approximately similar to a sample from a normal distribution. In mathematical statistics confidence intervals are usually derived in some asymptotic context,i.e. the distribution of a sample approaches a normal one as the sample size $N$ goes to infinity; then for large $N$ it is possible to approximate the sample distribution
 by a normal one and find an approximate confidence interval. This is done by taking the mean and variance of the sample, fitting a normal distribution to it and calculating for it a confidence interval.
 In more applied statistics such intervals are computed if only the form of the sample looks similar to a normal
 one by heuristic considerations; for example visual inspection or if the third and fourth moments are approximately close to those of a normal distribution. But in any case, an approach which just uses the mean and the standard deviation (coefficient of variation) of a sample for computing confidence intervals without any further considerations is quite problematic (see \citet{Wilcox(2009)}).
 
 Here now for the  example in section $3$, which is taken from chapter 5 of \citet{Au/Wang(2014)},  $500$ runs of the SuS algorithm were made and then the histogram of the estimators for $\Pr(g(u_1,u_2)<0)$ was plotted.
 For this histogram then a qq-plot (see e.g. \citet{NISTstat(2012)}) was made to examine if it follows approximately a normal distribution. 
\begin{figure}[htb]
 	\begin{center}
 		\includegraphics[width=8cm,height=8cm]{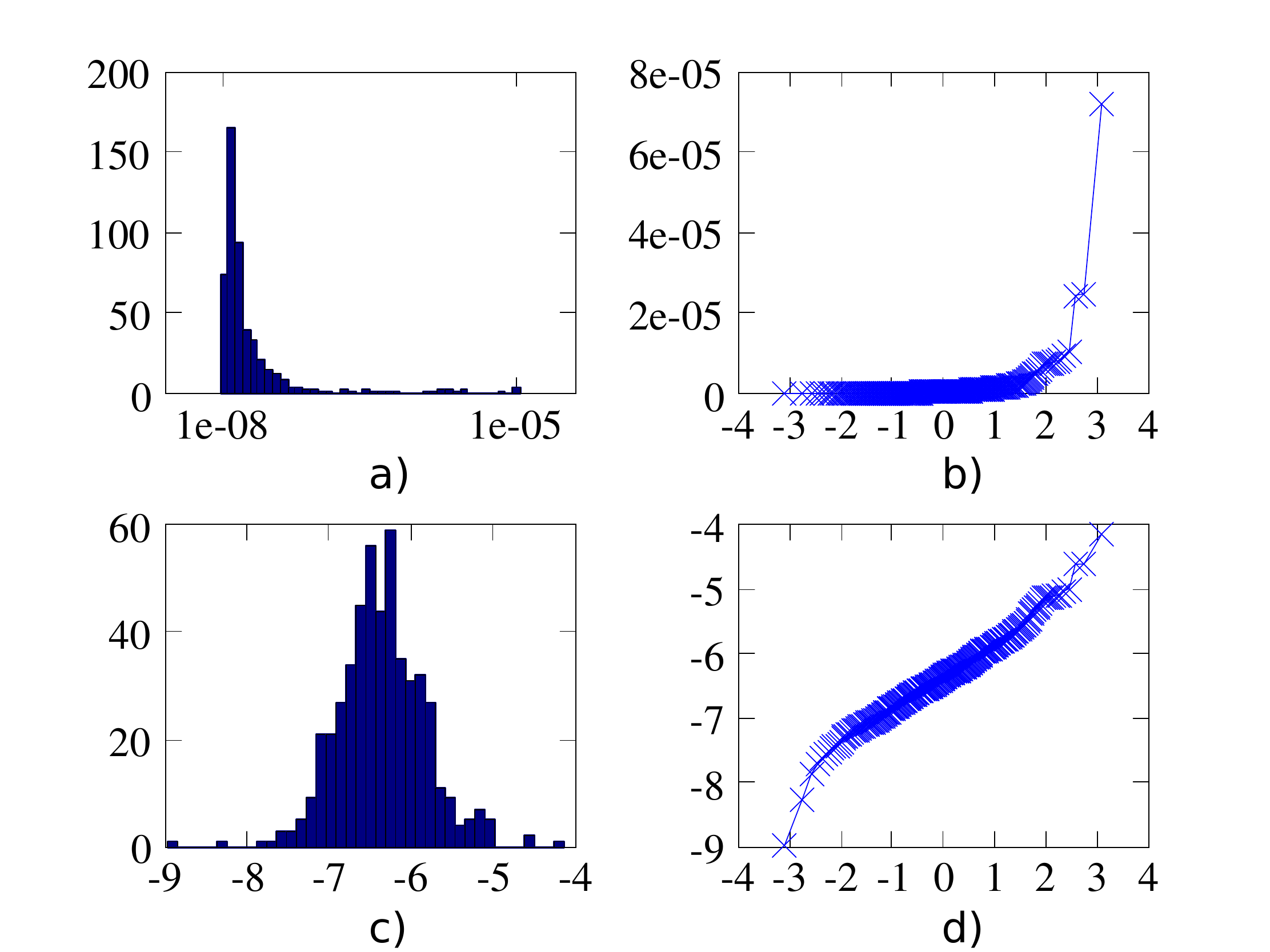}
 		\caption{a) histogram of data, b) qqplot of data, c) histogram of decimal logarithms of data, d) qqplot of logarithms}
 		\label{fig:exampleqq}
 	\end{center}
\end{figure}
 One can see from the qq-plot in figure \ref{fig:exampleqq}\ b) that the histogram is quite different from a normal distribution.  The recommendation in \citet{NISTstat(2012)} is  to use for such skewed data a lognormal or Weibull distribution. If the logarithms are taken, their distribution is approximately near a normal curve.
 Then it is meaningful to construct a confidence interval based on the standard deviation of the transformed data, whereas a confidence interval based on the first two moments of the untransformed data is not meaningful.
\section{Examples}
All the   examples  were calculated with the SuS algorithm given in \citet{Li/Cao(2016)}.
As parameters were taken $500$ samples per step, an
acceptance probability of $0.1$ and a chain length of ten. This  setup was used in \citet{Au/Wang(2014)} for estimates of the context of two-dimensional domains.

Always the left plot in a figure  with label a) shows the contour plot of the limit state curve $g(u_1,u_2)=0$ in blue  and the right plot with label b) the graph of the LSF, i.e. the surface $\{(u_1,u_2,g(u_1,u_2))\}$ as a  blue mesh.
Also the plane $z=0$ is shown as a a tranlucent grey plane.
The SuS sample points in the diagrams are marked by  green points. To improve the readability of the diagrams, only every tenth sample point is shown. Further in the diagrams on the left the design points for the LSF are shown as red circles.
\subsection{Piecewise Linear Functions}
The simplest cases where SuS approximations might become misleading are piecewise linear functions.
If the decrease velocity of such functions changes in a non-monotonic way in different directions,
the search path of SuS can be led away from the design point.
Consider here  two piecewise linear functions $g_1$ and $g_2$ defined by 
\ba
g_1(u_1,u_2)&=&
\begin{cases}
4- u_1 &,  u_1 > 3.5 \\
0.85- 0.1\cdot u_1&, u_1 \leq 3.5
\end{cases}
\\
  g_2(u_1,u_2)&=&
  \begin{cases}
  	0.5-0.1\cdot  u_2 &,  u_2 > 2 \\
  	2.3-  u_2&, u_1 \leq 2
  \end{cases}
 \ea
\begin{figure}[hbt]
 	\begin{center}
 		\includegraphics[width=8cm,height=5cm]{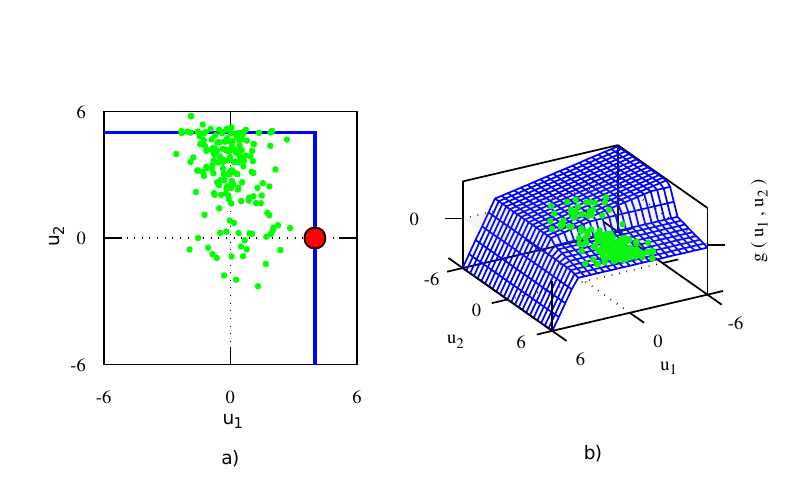}
 		\caption{}
 		\label{fig:piecelin}
 	\end{center}
\end{figure}
The series system defined by these two LSF's has the LSF $g^*(u_1,u_2)=\min(g_1,g_2)$.
The performance of the SuS method for this LSF  is shown in figure \ref{fig:piecelin}.
Due to the switching  decrease velocities of the two LSF's the SuS sample points move after these 
switches  happen into the wrong direction.
\subsection{Extrapolation}
Everybody is extrapolating almost all the time, in science and in  private live. 
So, extrapolation is a necessary and indispensable tool, but also it can be  dangerous  in reliability. Well known examples here are the Challenger disaster and the railway accident near  Eschede.
In civil engineering  stress-strain curves give a simple example; in the first part usually linear, they become non-linear
after the  elastic limit depending on the material.
So, extrapolating from the linear relationship at the lower stress rates would give wrong  values.

Consider now an example in reliability. In many  cases it seems to be  reasonable to assume that the tail distributions of random variables are different from their distribution in the central part (\citet{Maes(1994b)}, \citet{Acar/Ramu(2014)}). The next example shows the problem of extrapolation with SuS. Let be  given the  LSF
\ba
g(u_1,u_2)=0.1\cdot (52- 1.5\cdot u_1^2- u_2^2).
\ea
\begin{figure}[hbt]
	\begin{center}
		\includegraphics[width=8cm,height=5cm]{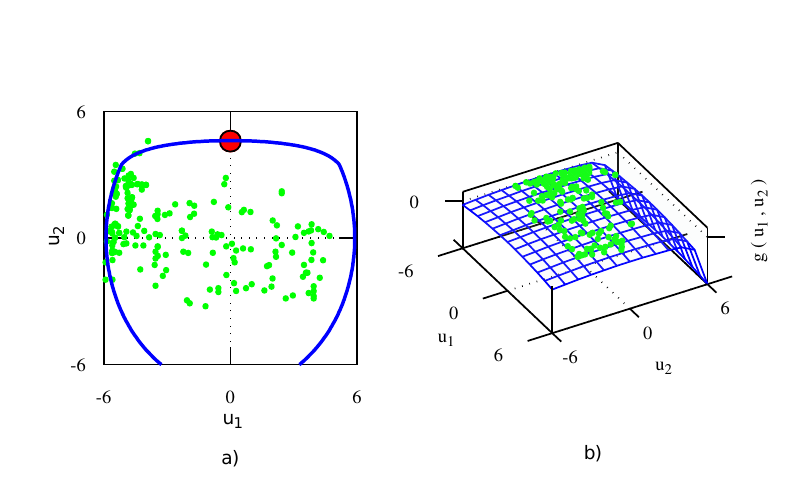}
		\caption{Distribution with Pareto tail ( solid curve limit state curve)}
		\label{fig:pareto}
		\end{center}
	\end{figure}
In the original space the first random variable has a standard normal distribution and the second  a standard normal distribution up to $a=3.5$ and  for the upper tail a Pareto distribution is fitted. These are then transformed into the standard normal space.
More formally written, in the original space there are two independent rv's $X_1$ and $X_2$, where $X_1$ has a standard normal distribution and $X_2$ has the following CDF $F_2(x_2)$:
\ba
F_2(x_2)&=&
\begin{cases}
	\Phi(x_2) &,  x_2 \leq  3.5 \\
	1-x_2^c&, x_2 > 3.5
\end{cases}
\ea
with $ c=\log(\Phi(-3.5))/\log(3.5)$.

So the upper tail of this rv has a Pareto distribution.
As one can see in figure~\ref{fig:pareto} the SuS samples move towards the local distance minima of the limit state surface on the horizontal axis, whereas the global minimum lies on the positive vertical axis.
So following the distributional form around the origin leads the algorithm in the wrong direction, since for larger values of $u_2$ the LSF decreases then much faster.
\subsection{Invariance}
An important reason that the Hasofer-Lind  index was adopted as a measure for reliability is its invariance under reformulations or re-parametrizations of the underlying  reliability problem (see \citet{Hasofer(1974)}).
Also the convergence proofs for the beta point search algorithms do not depend on the specific form of the LSF. But clearly one has to start the search algorithm from different points to obtain all global minimal distance points, i.e. beta points.
\begin{figure}[htb]
 	\begin{center}
 		\includegraphics[width=9cm, height=5cm]{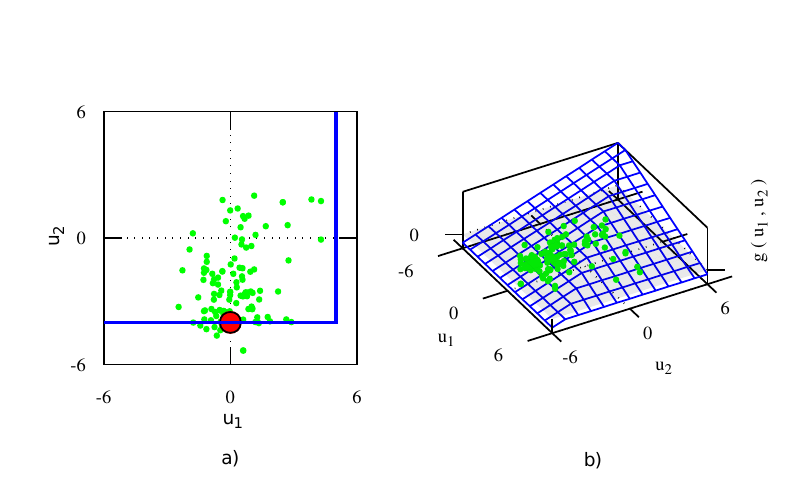}
 		\caption{Series system defined by LSF in eq.~(\ref{linser})}
 		\label{fig:lin}
 	\end{center}
\end{figure}  
 Consider now  a series system consisting of  two independent components, so failure occurs if at least one  fails.
 The first component fails if $u_1>5$ and the second component if $u_2<-4$.  
 Now, this limit state surface  can be the zero set  of different LSF's. 
 For example, one has 
 \ba\label{linser}
 g(u_1,u_2) =\min \begin{cases}
 	5-u_1\\
 	4+u_2
 \end{cases}
\ea 
 Here both LSF's are linear,  with  SuS  one obtains as expected an estimate for the asymptotic failure probability approximation $\widehat{P(F)}\approx \Phi(-4)\approx 3.17\cdot 10^{-5}$.
 The performance of the SuS method is shown in figure~\ref{fig:lin}.

 Assume now that  the LSF for the second random variable is given not by a linear but by a logistic function in the form:
 \ba
  g_2^*(u_2)=\frac{1}{1+\exp(-2( u_2+4))}-0.5
   \ea
   Then the LSF  $g^*(u_1,u_2)$ given by 
   \ba\label{nonser}
   g^*(u_1,u_2)=\min\begin{cases}
   	5-u_1\\
   	\frac{1}{1+\exp(-2( u_2+4))}-0.5
   \end{cases}
   \ea
    defines the same limit surface as before, but the shape of the LSF is different and the contour lines of these  functions are different in the safe and unsafe  domain, both LSF's have only the contour of zero level set in common.
%
\begin{figure}[htb]
   	\begin{center}
   		\includegraphics[width=8cm,height=5cm]{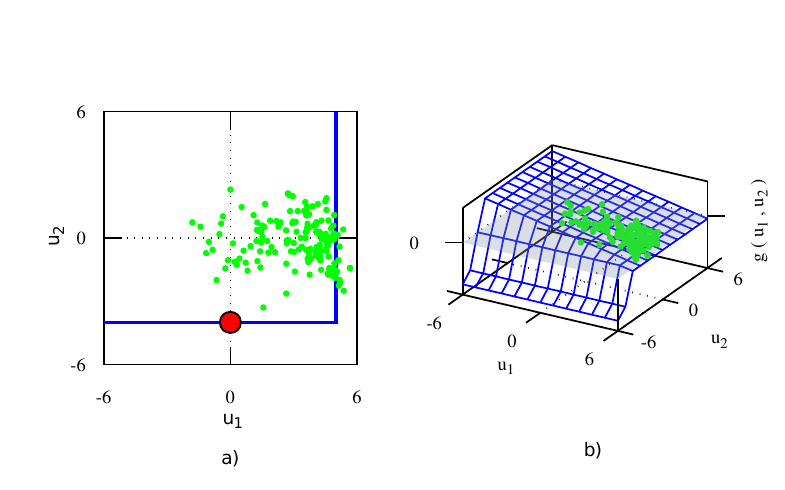}
   		\caption{Series system  defined by LSF in eq.~(\ref{nonser})}
   		\label{fig:logistic}
   	\end{center}
\end{figure}  
 Here, with the LSF  defined in eq.~\ref{nonser}, the points in SuS  converge towards the point $(5,0)$ and one gets as probability estimate a value  of $\Phi(-5)\approx 2.87\cdot 10^{-7}$ whereas the true failure probability
 is approximately equal to $\Phi(-4)\approx 
  3.17\cdot 10^{-5}$ as shown in figure \ref{fig:logistic}.  So, here the different forms of the LSF's influence the result of the method. The reason is that the structure of the LSF in the neighborhood of the origin is different from its form near the limit state surface.
 
 The same limit state surface can be described by a plethora of different LSF's. Their specific forms will influence the behavior of the SuS algorithm. Especially for more complicated LSF's for series or parallel systems it might be useful to clear inasmuch this can create convergence problems or lead to incorrect results.
 Certainly there will be cases where the result will not depend on the changing structures of the LSF's, but as the example above shows, it would be overoptimistic to assume that this is generally so.
%
\subsection{Changing Topological Structure of Domains}
Another case is when  topological structures of the failure domains changes, for example if its genus changes.
Assume that an  LSF is given by a metaball function (\citet{Metaball(2016)}):
\ba
g(u_1,u_2)=d - \sum_{j=1}^k\frac{c_j}{(u_1-a_j)^2 +(u_2-b_j)^2+1}
\ea
with  $k$ a natural number and $a_j,\ b_j$ and $ c_j$  real numbers.
If the parameter $d$ changes, the contours given by $g(u_1,u_2)=0$ take on different shapes.
\begin{figure}[htb]
	\begin{center}
		\includegraphics[width= 8cm,height=4.5cm]{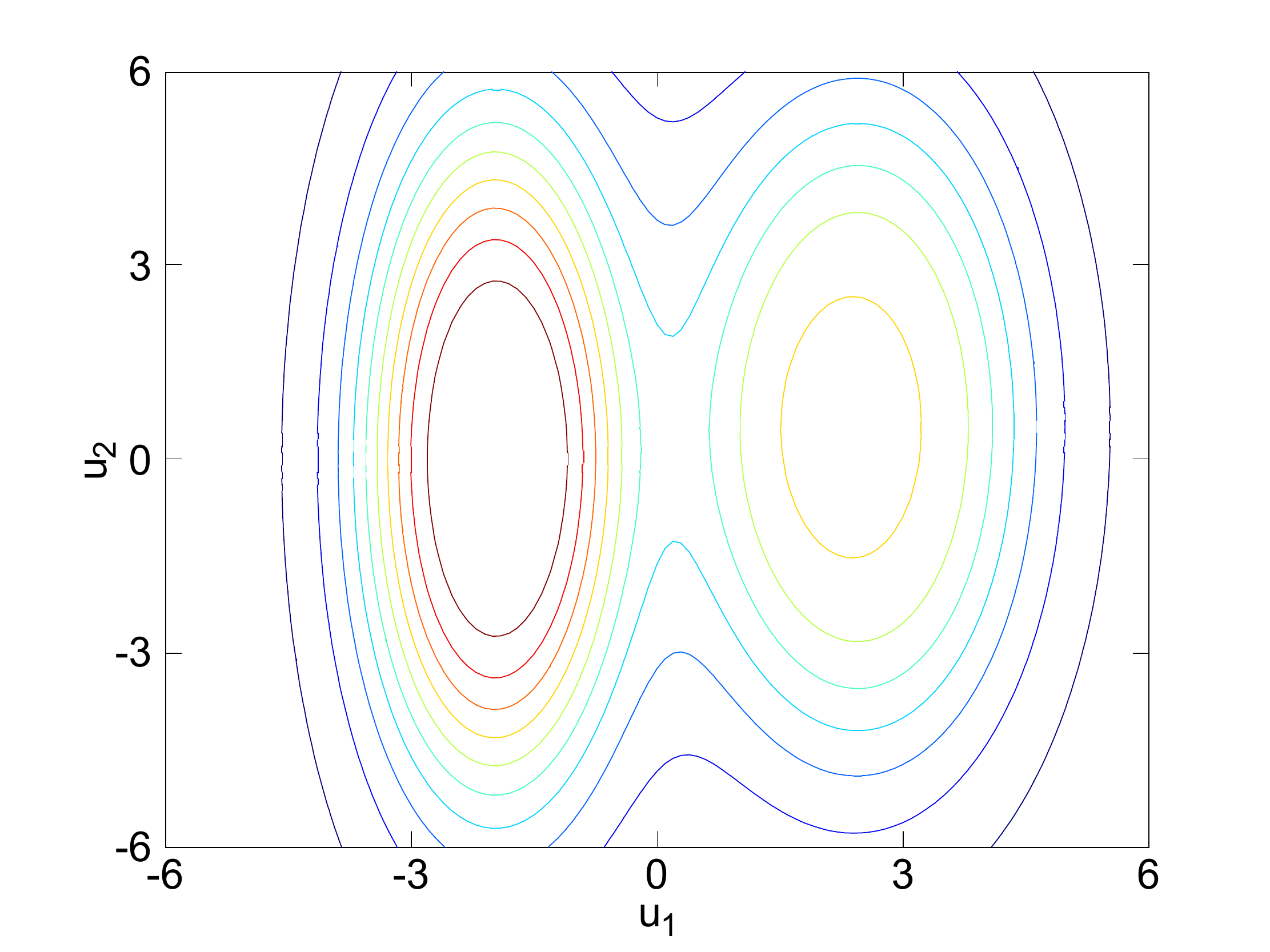}
		\caption{Example for LSF's created by eq.~\ref{multiball}  }
		\label{fig:genus}
	\end{center}
\end{figure}
 As example consider now a simple metaball function defined by: 
  \ba\label{multiball}
  g(u_1,u_2)&=&
  {{30}\over{\left({{4\,\left(u_1+2\right)^2}\over{9}}+{{u_2^2}\over{25
  			}}\right)^2+1}}\\ &+&{{20}\over{\left({{\left(u_1-2.5\right)^2}\over{4}}+
  			{{\left(u_2-0.5\right)^2}\over{25}}\right)^2+1}}-5
  	\ea
\begin{figure}[htb]
  		\begin{center}
  			\includegraphics[width=8cm,height=5cm]{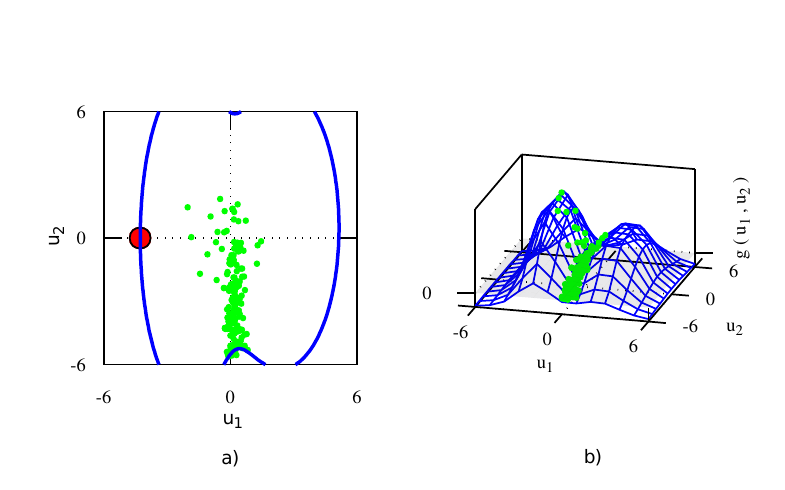}
  			\caption{SuS for the LSF's in fig.~\ref{fig:genusex} }
  			\label{fig:genusex}
  		\end{center}
\end{figure}
  	For decreasing values towards zero  that the safe domain consists first of two elliptic regions which then merge to one region.
For larger values of the parameter the failure domain has topological genus two which then changes to one. To formulate it more sloppy, first there are two holes in it and then only one. This can be seen in figure \ref{fig:genus}.

In figure \ref{fig:genusex} the SuS results for one run for this  example are shown. 
The sudden change in the topological structure creates difficulties and the sample points move in the wrong direction.
\subsection{Rotationally Unsymmetric LSF's}
The  von Mises distribution (see e.g. \citet{Forbes(2011)}) is defined on the unit circle and has there the PDF:
\ba
f(\varphi)=\exp(\kappa\cos(\varphi-\mu))/(2\pi I_0(\kappa)),\ 0\leq \varphi < 2\pi
\ea
The parameters $\mu$ ($0\leq \mu <2\pi$) and $1/\kappa$ ($0<\kappa$) are location resp. dispersion measures and  $I_0(.)$ is the modified Bessel function  of order $0$.
Using a LSF proportional to a mixture of two non-normalized densities  and for the distance to the origin functions with varying decrease velocities
\ba\label{cramer}\nonumber
\hspace{-1cm}g(r,\varphi)=0.19- 0.0055\cdot(\Phi(r-0.5)\cdot\exp(4\cdot \cos(\varphi))\\
 - 12\cdot (\Phi( 0.004r)-0.5)\cdot \exp(\cos(\varphi-\pi))
\ea
\begin{figure}[htb]
	\begin{center}
		\includegraphics[width=8cm,height=5cm]{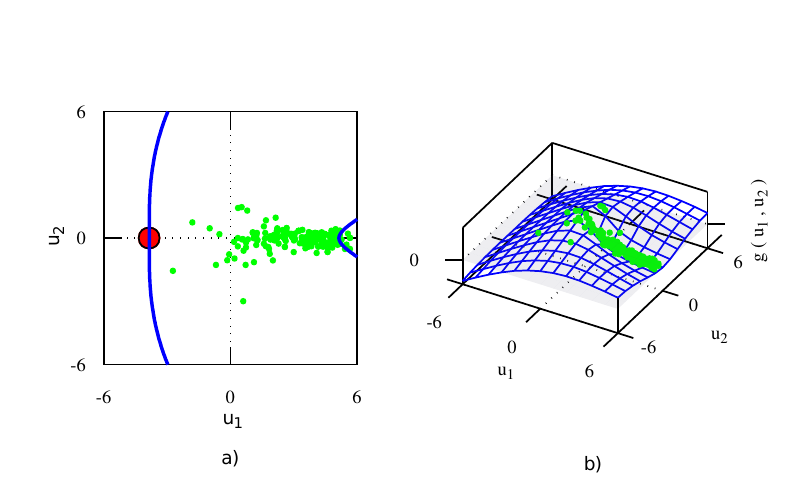}
		\caption{SuS for the LSF's in fig.~\ref{fig:cramer} }
		\label{fig:cramer}
	\end{center}
\end{figure}
This LSF shows different behaviors in different directions from the origin.
Since the LSF decreases in one direction, $\varphi=0$, at the beginning faster and then the decrease slows down,
whereas in the direction $\varphi=\pi$ it is viceversa, the algorithm moves into the wrong direction.
\subsection{Several Beta Points}
\begin{figure*}[h!tb]
	\centering
	\begin{subfigure}[b]{0.5\textwidth}
		\centering
		\includegraphics[width=8cm,height=5cm]{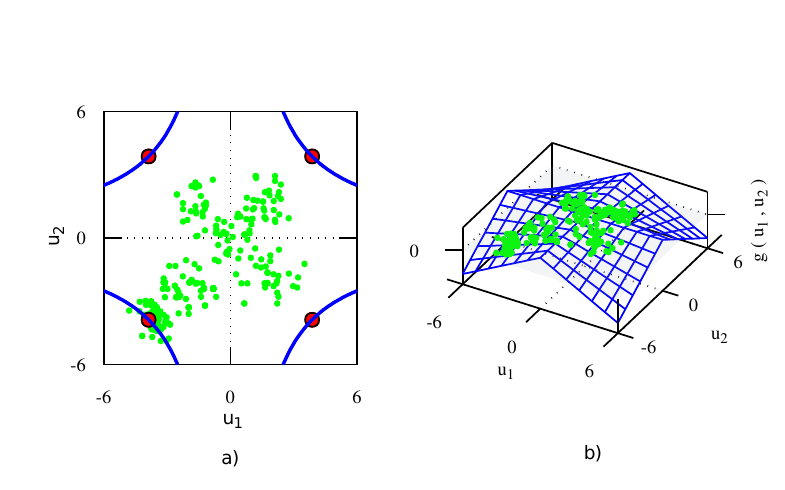}
		\caption{SuS detects one beta point}
		\label{fig:one}
	\end{subfigure}%
	\hfill
	\begin{subfigure}[b]{0.5\textwidth}
		\centering
		\includegraphics[width=8cm,height=5cm]{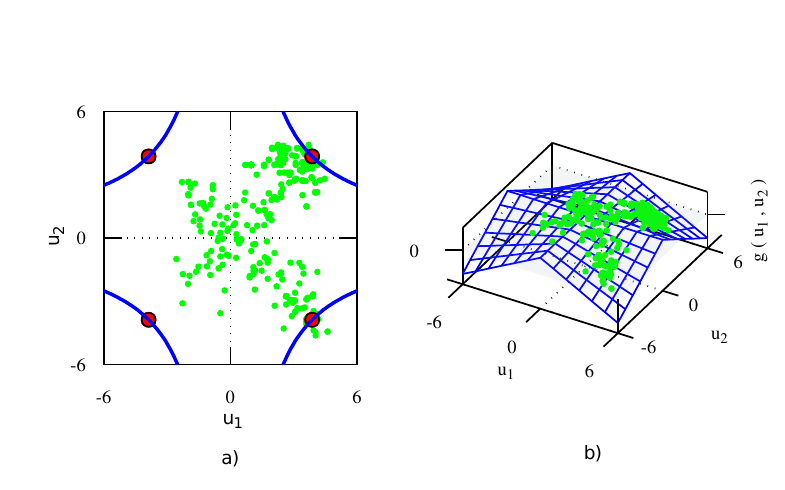}
		\caption{ SuS detects two beta points}
		\label{fig:two}
	\end{subfigure}
	\vskip\baselineskip
	\centering
	\begin{subfigure}[b]{0.5\textwidth}
		\centering
		\includegraphics[width=8cm,height=5cm]{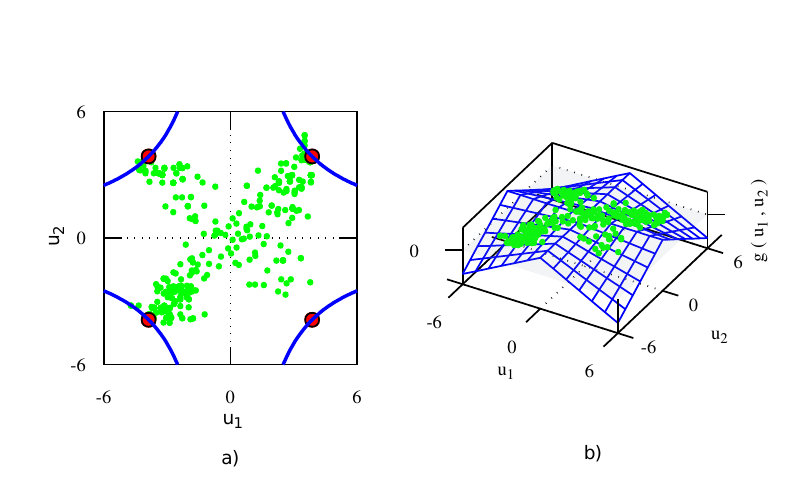}
		\caption{SuS detects three beta points}
		\label{fig:three}
	\end{subfigure}%
	\hfill
	\begin{subfigure}[b]{0.5\textwidth}
		\centering
		\includegraphics[width=8cm,height=5cm]{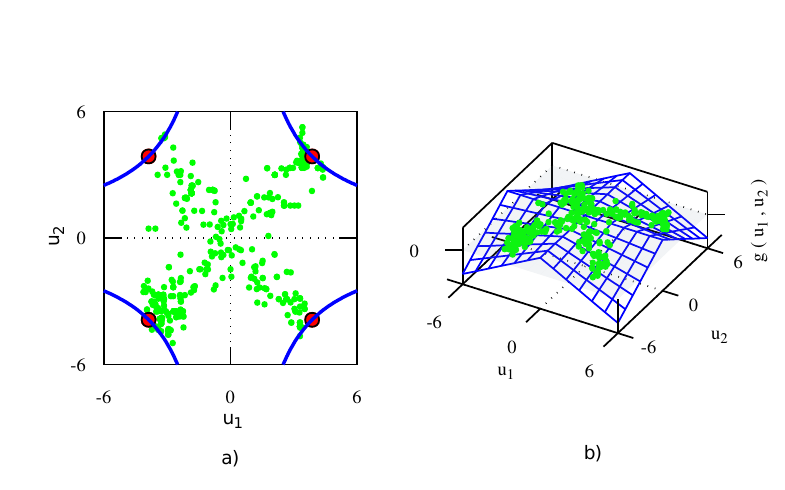}
		\caption{ SuS detects four beta points}
		\label{fig:four}
	\end{subfigure}
	\caption{SuS for the LSF $g(u_1,u_2)=15-|u_1\cdot u_2|$}\label{fig:runsus1}	
\end{figure*}
Consider now a slightly more complicated version of the example studied in the second paragraph.
Let  the LSF be 
\ba
g_{\beta}(u_1,u_2)= \beta^2/2 - |u_1\cdot u_2|.
\ea
Due to the symmetry of the LSF   there are four beta points.
In a FORM/SORM analysis one obtains  using the results found in  the following asymptotic approximation
  one has for the failure probability 
$\Pr(g_{\beta}(u_1,u_2)<0)\sim 2\sqrt{2}\cdot\Phi(-\beta) ,\ \beta\to\infty$.
This is obtained by just doubling the result in eq.~(\ref{exact}).
Here clearly in a SORM analysis the beta point search algorithms have to be started several times to find all beta points.

If  this problem is examined now with SuS  the possible  outcomes of runs are shown in figure~\ref{fig:runsus1}.  
In fifty runs of SuS in one case only one beta point was detected, in 11  two, in 29  three  and only in nine cases all four were found.
This might lead to a systematic  underestimation of the failure probability when not  all beta points are found.
If now several runs are combined, there will still be a bias, the failure probability will be underestimated.
It is  unclear to the author  how to get a good estimator of the failure probability here  without making some sort of geometric analysis similar to FORM/SORM.
 \subsection{The black swan example from Au/Wang}
\begin{figure}[htb]
 	\begin{center}
 		\includegraphics[width=8cm,height=5cm]{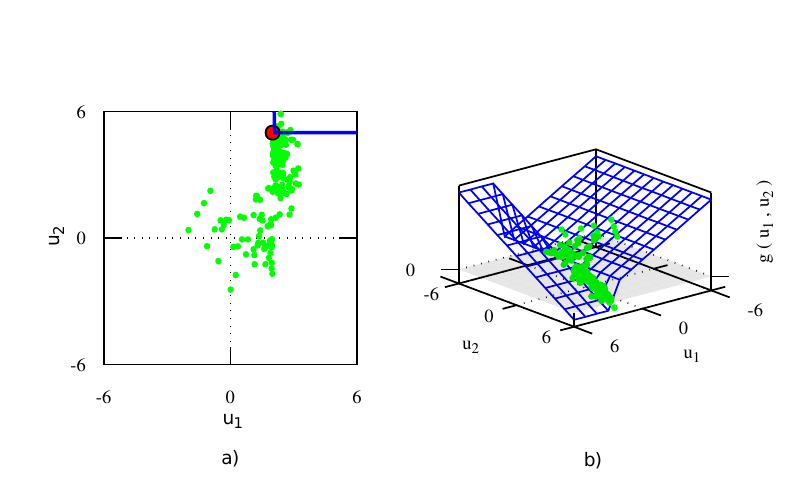}
 		\caption{SuS for the LSF from chapter 5, p.~197 in \citet{Au/Wang(2014)}  }
 		\label{fig:auwang}
 	\end{center}
\end{figure}
 In \citet{Au/Wang(2014)} in chapter 5 there is an example to illustrate a case where SuS has difficulties to reach the failure domain. This example is shown here. One can see that the problem in this case is essentially different from the other problems shown before. The data points move into the right direction, the difficulty is only if the dispersion of the points is large enough to reach the failure region. Au/Wang propose to analyze this specific problem  by looking at the cumulative distributions curves if there are any doglegs there. But the other examples treated before require an understanding of the geometry of the failure domains which in general cannot be found from the cumulative distribution functions as least as far the author knows.
\section{Conclusions}
The subset simulation  method --- in the right circumstances --- can give    good failure probability estimates, but it has its   limitations as  was tried to explain here. These problems were illustrated by  relatively simple two-dimensional  examples to provide an intuitive idea of the possible shortcomings. This should aid to a better understanding of SuS. Further it
should guide to clarify where the restrictions of SuS are precisely and how to detect those for a given problem.
Unfortunately, the bilk of SuS literature is about increasing the efficiency of the method and not about understanding how it works or not; as is argued in \citet{Hooker(1995)}, research should be concentrated more about this. Since certainly not so efficient, but correct algorithms are more desirable than very high efficient ones which give wrong results. Further extensive Monte Carlo studies, which are recommended in \citet{Bartz(2010)} for clarifying the performance of algorithms if no general theoretical results can be derived, are lacking.

The examples demonstrate that the  points chosen by SuS usually move in the directions of steepest descent of the LSF's near the origin, but  latter changes in the descent speed of the LSF's and changes in the topological structure of the failure domains may lead the SuS estimators  in wrong directions.
A   disadvantage of SuS for detecting more complex structures in limit state surfaces seems to be in the opinion of the author this underlying idea  of extrapolating  from failure domains nearer to the origin towards the original limit state surface which is far away from the origin. Here it is taken for granted  that the structure does not change essentially during the extrapolation movement. It appears not too easy to justify this assumption in a general way.
%

 So the SuS proponents have the problem that they claim that their method is a progress above FORM/SORM, but with their method they cannot gain information about the limit surface. This is  even declared  as a feature of SuS (\citet{subsetsampling(2016)}):\\
 \textit{Subset Simulation takes the relationship between the (input) random variables and the (output) response quantity of interest as a 'black-box'.} 

Grave  problems --- in the opinion of the author --- result from this attempt in  SuS concept to avoid all   geometric concepts used in FORM/SORM. The effect is that information which can gained by modeling the geometric structure of the limit state surface is not used; this can  lead to  slowing down the procedure or to not finding correct estimates. In the examples here it is possible to detect by visual inspection  this, but in higher dimensions it seems to be possible only by an analysis of the structure of the limit state surface.

The second last example treats the case of a failure domain having several disjoint subsets which results in several design points. Here the problem is to identify all these sets/points, which does not succeed always.
As Rackwitz  said in \citet{Rackwitz(2001)}: 

{\textit{Unfortunately, the presence of multiple critical points is not as infrequent as one might wish.
		They can occur in standard space as well as in original space.\ldots 
		Usually, the most difficult part is to know whether such a problem exists at all.}}
 
But in the SuS concept this problem does not exist at all, since the structure of the LSF is seen as a black box. Now, look at possible LSF structures and test examples for SuS, where one plays the \textit{advocatus diaboli} by setting up difficult tests and the other tries to solve them using SuS.
The problem seems a the first glance to be  similar to the old German fairy tale about the hare and the hedgehog where here both can play whatever part they want. Even ignoring the problem of multiple beta points, by increasing the number of samples and runs  one can always find a good estimator for the failure probability with SuS, but there can be constructed always a more complex problem for which still more simulation effort is necessary. Finally one reaches the extent of a full blown Monte Carlo analysis.
 But the analogy limps, since the real problem is that the user of SuS does not see --- if he applies only SuS --- if his results are not correct and need improvement. So theoretically, if he would know the correct result in advance he could improve the SuS estimates by increasing sample size and/or run length, but this is usually never the  case in more realistic problem settings. Since from the results of SuS one obtains no information about possible wrong directions the algorithm has taken, all problems are strapped on the same Procrustes bed of the method and then the results are registered.

   If one assumes that the LSF is given in black box form, not explicitly, then making a FORM/SORM analysis of the LSF's, it  would be possible to extract all this information about the design points and the curvatures of the limit state surface from the results of this analysis. Instead if a SuS analysis is made, the
 result would be the histogram of  complimentary cumulative distribution function estimating the distribution of the LSF $g(\uu)$.

 

 Compared with FORM/SORM, where the identification of the relevant beta points is not too time consuming, the advantages  of the SuS method seem to be  lost in such circumstances. Certainly as said by using more samples for all the problems above the correct solution can be  found. But as already said it remains unclear when one should increase the sample size and/or the number of runs. And might it be possible by improving the method to check if the probability contributions from all relevant parts of the failure domain have been found? If SuS is seen as stand-alone method and not as an add-on for FORM/SORM, it appears difficult to achieve all these goals. 
So there are some points in SuS concept where a clarification and more careful elaboration of the approach would be very desirable and the development of further refinements would be helpful and important.
 
In the moment, as it stands, the SuS approach seems to be a step backwards concerning the identification of the structure of the failure domain, since this topic is ignored there. This does not necessarily mean that this cannot be resolved by a modified  SuS, but if  it is possible it is really time  to do this. 

The SuS approach is focused on the computation of failure probabilities for a given probabilistic model.
Such has been one of the paradigms of structural reliability for the last fifty years. But the problems studied here are changing and it might be good to reconsider the basic concepts a little bit. Now, when more and more high dimensional structures are studied, where no intuitive understanding of the limit states is possible anymore,
is this really what one wants? Everybody knows that the numbers are wrong anyway, so is having an algorithm spitting out numbers/probabilities in fact that what the analyst or his client wants and needs?

In \citet{Breitung(2017a)} there is a first attempt to discuss if not a shift of paradigms from the computation of probabilities to the detection of structures might be meaningful. This would lead to a change from seeing the basic problem of structural  reliability in calculating probabilities towards  investigating the structure of limit state surfaces and failure domains.

To finish there are some problems in the SuS approach to structural reliability calculations. These gaps need to be filled to keep this concept as a serious contender in this field, otherwise it might be helpful to clarify the limitations of this method by Monte Carlo studies and to avoid to apply SuS in problems beyond these limits.  Until either the first or the second goal has been achieved, SuS should be applied only --- at least in the opinion of the author --- if the correctness of the procedure can be verified by some external method not based on SuS.
\section*{Acknowledgement}
The author thanks Prof. Hong-Shuang Li from the Nanjing University of Aeronautics \& Astronautics for providing him a preprint of his article and  explanations about the algorithm there. And  he thanks Dipl.-Ing. Ch. Gasser from the TU Vienna for some calculations and discussions about SuS.
\Urlmuskip=0mu plus 1mu\relax
{

\end{document}